\RequirePackage{fix-cm}
\makeatletter
\def\cl@chapter{}
\makeatother

\documentclass[twocolumn]{svjour3}
\smartqed

\usepackage[utf8]{inputenc}
\usepackage[T1]{fontenc}
\usepackage[unicode]{hyperref}
\usepackage[english]{babel}
\usepackage{url}

\usepackage{amsmath,amssymb,amsfonts,bm}
\usepackage{graphicx}
\usepackage{algorithm,algorithmic}
\usepackage{tcolorbox}
\usepackage{wrapfig}
\usepackage{subcaption}
\usepackage{multirow}
\usepackage{tabularx}
\usepackage{float}
\usepackage{enumitem}
\usepackage{xcolor}

\usepackage[nameinlink,capitalize]{cleveref}

\newcommand{\etal}{\textit{et~al.}}

\newcommand{\pos}{\mathbf x}

\newcommand{\eye}{\mathbf e}

\DeclareMathOperator{\maskValue}{maskValue}
\newcommand{\norm}[1]{\left\lVert#1\right\rVert}

\newsavebox{\largestimage}

\crefformat{footnote}{#2\footnotemark[#1]#3}
\crefname{section}{Sec.}{Sec.}

\journalname{arXiv} 
\begin{document}

\title{Volume Conductor: Interactive Visibility Management for Crowded Volumes}

\author{Žiga Lesar$^1$ \and Ruwayda Alharbi$^2$ \and Ciril Bohak$^{1,2}$ \and Ondřej Strnad$^2$ \and Christoph Heinzl$^3$ \and Matija Marolt$^1$ \and Ivan Viola$^2$}
\institute{
    $^1$Faculty of Computer and Information Science,\\
    University of Ljubljana, Večna pot 113,\\
    1000 Ljubljana, Slovenia\\
    $^2$Visual Computing Center,\\
    King Abdullah University of Science and Technology (KAUST),
    Thuwal 23955-6900, Kingdom of Saudi Arabia\\
    $^3$University of Applied Sciences Upper Austria,\\
    Roseggerstraße 15, 4600 Wels, Austria
}
\date{ }

\maketitle

\begin{figure*}
    \centering
    \includegraphics[width=\linewidth]{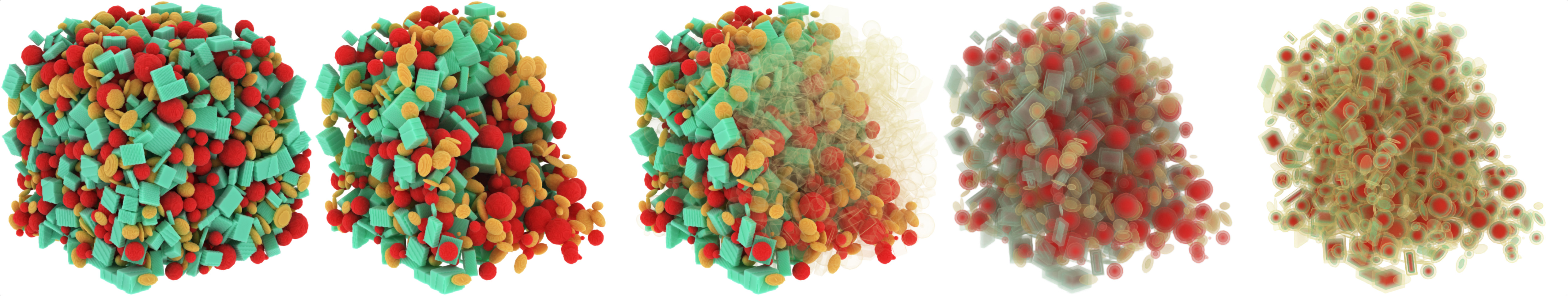}
    \caption{Features of the volume conductor. From left to right: instance grouping and colorization, sparsification, ghosting, blending with raw data, and opacity transfer from sparsification to raw data.}
    \label{fig:teaser}
\end{figure*}

\begin{abstract}
We present a novel smart visibility system for visualizing crowded volumetric data containing many object instances. The presented approach allows users to form groups of objects through membership predicates and to individually control the visibility of the instances in each group. Unlike previous smart visibility approaches, our approach controls the visibility on a per-instance basis and decides which instances are displayed or hidden based on the membership predicates and the current view. Thus, cluttered and dense volumes that are notoriously difficult to explore effectively are automatically sparsified so that the essential information is extracted and presented to the user. The proposed system is generic and can be easily integrated into existing volume rendering applications and applied to many different domains. We demonstrate the use of the volume conductor for visualizing fiber-reinforced polymers and intracellular organelle structures.

\end{abstract}

\section{Introduction}
\label{sec:introduction}

Surface rendering methods used in modern 3D applications typically focus on object boundaries and thus provide insufficient insight into volumetric datasets found in many branches of science. Such volumetric datasets can be visualized with direct volume rendering (DVR) methods, which are often coupled with \emph{visibility management} techniques to reveal or emphasize specific structures or regions of interest. Examples include transfer function specification, planar reformation, clipping geometry, cutaway views, exploded views, and spatial deformations.

In many cases, a volume is densely populated with numerous instances of structures that occlude each other, and the absence of visibility management results in an uninterpretable and cluttered image. We call such volumes \emph{crowded volumes}. Examples include scans of polycrystalline materials, fiber-reinforced polymers, and intracellular biological structures. For such data, existing volume rendering methods are not suitable because the amount and density of the instances are so high that occlusion impedes the spatial perception of patterns and distributions in the data. Usually, no single instance is of particular interest, but rather their distribution inside the volume; therefore, a suitable visibility management strategy is to \emph{sparsify} the volume by removing or fading some instances. The instances are often characterized by additional attributes, such as volume, surface area, orientation, position, and others, and the users are typically interested in investigating the spatial distribution of the instances with specific attributes.

For the visualization of such volumes, we propose the \emph{volume conductor}, a novel interactive exploration and visibility management system for crowded volumes. 
With the volume conductor, the user directs the color and visibility of groups of instances to obtain the desired visualization of a crowded volume. The volume conductor provides an easy-to-use user interface for forming groups of instances and managing their visibility. The user can directly adjust the sparsification of each group through manipulation of scented sliders~\cite{Willett2007}. The sparsification routine automatically determines which instances are visually suppressed and which remain visible, which is a significantly different process compared to transfer-function-based visibility management, where the visibility is implicitly controlled. The user can rapidly achieve the desired visualization by organizing instances into a hierarchy according to their attributes. We also present a technique for combining the resulting sparsified volume with the raw data volume, which enables integration of the volume conductor with existing DVR techniques. An overview of these features is presented in \cref{fig:teaser}. The sytem is generic and can be applied to many domains, which we demonstrate on three use cases.

We emphasize the following original contributions of our work:
\begin{itemize}[noitemsep]
    \item rendering-method-independent interactive visibility management system for visualizing crowded volumes;
    \item combined rendering of sparsified segmented volumes and raw data volume for providing the context of displayed instances within the raw data;
    \item procedural generation of GPU shader code based on user-defined hierarchically organized instance attributes;
    \item domain-expert-defined use cases showcasing the benefits of the proposed system;
\end{itemize}

\section{Related Work}
\label{sec:related-work}


A volume is a dense representation exhibiting a high degree of mutual occlusion among the contained structures. Visiblity in volumes is usually directed with transfer functions~\cite{Ljung2016}. Traditionally, transfer functions map the intensities and their gradients to optical properties without awareness of the individual instances, and as such do not solve the occlusion problem. To prevent the occlusion of important information and provide enough context for better spatial comprehension, \emph{smart visibility management} techniques are integral part of volume visualization frameworks. Standard techniques include clipping planes, cutaway views, and automatic or interactive transfer function specification. Early approaches addressed the problem of the automatic generation of see-through technical illustrations~\cite{Diepstraten2002,Diepstraten2003} with view-dependent transparency techniques. Image-space depth sorting is used to generate semi-trans\-pa\-rent visualizations, combined with the existing rendering methods. Viola \etal~\cite{Viola2005} discussed how to visually expose essential parts of the data in volumetric renderings with an importance-driven feature enhancement technique enabling the automatic generation of cutaway and ghost views. Ament \etal~\cite{Ament2017} used selective illumination to automatically highlight important structures in a volume and make them visible from the camera. Moreover, Ropinski \etal~\cite{Ropinski2005} presented volumetric lenses for interactively enhancing interesting volume regions. Kubisch \etal~\cite{Kubisch2010} used breakaway views and ghost views, on a practical use case of tumor surgery planning. In addition, Chan \etal~\cite{Chan2008} explored spatial relations between the structures in a volume and suggested techniques for visualization.

While these techniques allow users to expose parts of the data in different ways, they do not offer precise control over which data should be exposed and to what degree. Moreover, none of these explicitly target crowded volumes and thus are not particularly useful for visualizing such data. In the best case, extensive preprocessing is necessary to prepare the data for visualization using a specific method. In contrast, our proposed approach offers interactive sparsification with minimal preprocessing and is independent of the rendering method.


\emph{Context-preserving methods} consider critical aspects of the visualization and retain them regardless of the view and data orientation. The magic volume lens~\cite{LujinWang2005} retains the context by deformation, instead of removal of structures, where the user selects the magnifying part. Krüger \etal~\cite{Kruger2006} presented a distance-based importance mapping of transparency to preserve the context surrounding the focused part of the data. A context-preserving method by Bruckner \etal~\cite{Bruckner2006} introduces an easily controlled context-preserving approach considering the shading intensity, gradient magnitude, eye distance, and previously accumulated opacity to reduce the opacity of less essential regions of the volume. This method is also adapted for one of the proposed sparsification functions, where it is used to control the importance of entire instances rather than individual voxels.

Context-preserving approaches work well on un\-crow\-ded data. However, for crowded volumes, their use alone is not enough to present and retain a global context around specific instances in a crowded environment. We complement the existing contributions by extending the voxel-based view of the data to an instance-based view through aggregation over the instances.


Streamline visualization is aiming towards automatic selection and rendering of the most representative instance in a crowded environment. Günther \etal~\cite{Gunther2013} addressed the problem by optimizing the streamline opacity, extended it for visualizing surfaces~\cite{Gunther2014}, sets of streamlines~\cite{Gunther2014a}, and a joint dataset with points, lines, and surfaces~\cite{Gunther2017}. Kanzler \etal~\cite{Kanzler2016} estimated a view-dependent visibility of the streamlines in screen space on a GPU roughly based on screen-space occupancy maps by Marchesin \etal~\cite{Marchesin2010} to handle line density control. Streamlines typically represent the properties of a continuous vector field, which is not the case for presented data, meaning we cannot use same sparsification approach. Moreover, these methods do not offer the user explicit control over the grouping, sparsification and emphasis of instances for which the volume conductor was explicitly designed.

Le Muzic \etal~\cite{LeMuzic2016} introduced \emph{visibility equalizers} as a tool to interactively sparsify mesoscopic biological data. Their data consist of molecular instances organized into a hierarchical representation. The visibility of these hierarchical groups is estimated in real time and can be adjusted through sliders. Visibility equalizers form the basis for a more general concept of the volume conductor presented in this paper. The approach supports only structurally identical hierarchically organized instances already and renders them through the instancing-enabled graphics pipeline. Instances are removed either randomly or based on the distance from a clipping primitive. For the volume conductor, this approach is generalized as a \emph{sparsification function}, and three different sparsification strategies are presented in this paper. A significant difference is that the volume conductor operates on instance properties where any property can be selected for sparsification, thus allowing complete flexibility for users to analyze the data according to any property of interest. For visibility equalizers, sparsification is hard-bound to the hierarchical scene arrangement, and no additional properties can be used for sparsification. In the volume conductor, this is enabled through the authoring environment, where properties are selected and organized into hierarchies. The reason for this is that the volume conductor additionally targets exploration and analytical scenarios, whereas visibility equalizers were primarily concerned with the communicative visualization intent.

In summary, although related to crowded environments, all of these methods fail to address crowded volumes and interactive exploration. Our contribution provides the user with complete control over instance grouping, sparsification, and rendering. While other approaches assume specific structures (e.g., streamlines or molecules), the volume conductor is domain-agnostic and does not pose any constraints on the structures and their properties. We support our claims with three use cases.

\section{Volume Conductor}
\label{sec:method}

\begin{figure*}[tb]
    \centering
    \includegraphics[width=\linewidth]{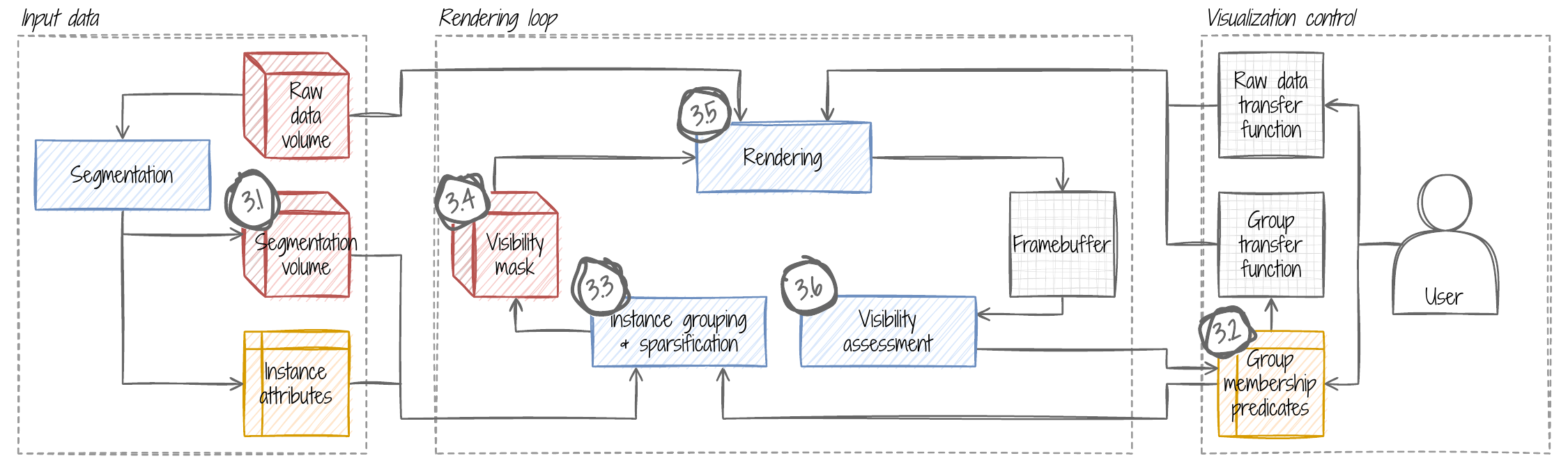}
    \caption{High-level overview of the volume conductor. Raw and segmented data pass through the grouping and sparsification procedure to generate the visibility mask, which is then rendered along with the raw data. Visibility is assessed from the rendered image and used to update the user interface. Circled numbers indicate section numbers, where that part of the volume conductor is described.}
    \label{fig:overview}
\end{figure*}

The volume conductor is an explicit visibility management technique for volumetric data, where the user forms groups of instances based on their attributes and controls the visibility of the instances with scented sliders~\cite{Willett2007} to communicate how many of them should be visible under the current transfer function and camera settings. This information is used by the sparsification procedure when determining the visibility of each instance. The architecture of our solution is displayed in \cref{fig:overview}. Two volumes are required at the input: the \emph{raw data volume} and \emph{segmentation volume}. In the latter, each voxel is assigned a numerical identifier corresponding to an individual instance or the background. As the segmentation process is not part of our pipeline, we do not propose any particular segmentation technique, and a suitable one can be chosen based on the given domain. Every instance may hold a set of additional attributes, such as length, volume, surface area, and orientation, supplied with the two volumes.

The user first forms groups of instances by specifying a list of \emph{group membership predicates}, which assigns each instance to a group. Depending on the user's visualization goals, the groups may contain instances of interest with specific attributes, instances that are less relevant and are to be sparsified, or even those that represent unwanted information, such as segmentation errors or noise. To facilitate this task, we developed an easy-to-use user interface for manipulation of group membership predicates (we discuss it in \cref{sec:instance-grouping}). After the groups are formed, the user assigns a color, opacity, and visibility ratio to each group. The colors may include transparency if the user wants a partially transparent group of instances. The visibility ratio is used in the sparsification procedure when determining the number of shown and hidden instances in that group. The user can choose between various sparsification functions to control which instances are affected. The sparsification procedure and various \emph{sparsification functions} are discussed in \cref{sec:instance-sparsification}. Based on the group membership predicates, the \emph{visibility mask} (a volume that encodes the visibility and group membership of every voxel) and a corresponding transfer function are generated for rendering. Their format and the generation process are described in \cref{sec:visibility-mask}. The visibility mask can be visualized with any existing volume rendering method, and we demonstrate this by integrating the volume conductor into directional occlusion shading (DOS)~\cite{Schott2009} and path tracing renderers. Finally, the user can \emph{blend the raw data} into the visualization to display information lost during segmentation or to enhance the raw data rendering with the capabilities of the volume conductor. The rendering process and blending are covered in \cref{sec:rendering-and-blending}. As the sparsification procedure does not consider occlusion between the instances when determining their visibility, we perform \emph{visibility assessment} by computing the actual visibility ratio of each group as observed from the camera. The ratio is calculated by rendering the segmentation volume into an ID frame buffer and counting the different IDs in each group. This information is fed back into the user interface, forming a feedback loop with the user and ensuring that the user settings are accurately reflected in the rendered image. This process is explained in more detail in \cref{sec:visibility-assessment}.

\subsection{Instance grouping}
\label{sec:instance-grouping}

A group membership predicate may be any Boolean expression that signals instance membership in a group based on its attributes. There is precisely one group for every predicate plus one additional group for the background. When computing the visibility mask, these predicates are traversed to distribute the instances into groups. If an instance satisfies several predicates, the one evaluated first defines the group of that instance; therefore, each instance belongs to exactly one group. If an instance does not satisfy any predicates, it is assigned to the background group.

Group membership predicates are defined in runtime by the user. To simplify the user interface, we decided to allow only predicates of a specific format easily represented in the user interface. Every predicate includes the following information that the user can adjust:
\begin{itemize}[noitemsep]
    \item the instance attribute that this predicate is based on,
    \item a series of ranges that the attribute value may fall into,
    \item the color of the group used when generating the group transfer function, and
    \item the visibility ratio between the number of shown and hidden instances in the group.
\end{itemize}

We provide two ways of defining the group membership predicates: sequential and hierarchical. A sequential set of predicates can be directly used to generate the visibility mask, whereas a hierarchical set must first be linearized. In this context, the child predicates of a single parent predicate form a disjunctive relation between themselves, whereas the nesting of the predicates signifies a conjunctive relation between the parent and child predicates. The evaluation of child predicates is short-circuited so that every child predicate is only evaluated if an instance also satisfies the corresponding parent predicate. Finding the first available group amounts to finding an admissible path in the hierarchy from a root predicate to a leaf predicate. This form of manipulation complies with a typical workflow to a large degree, where the user first broadly divides the instances into several groups and refines them with further subdivisions.

\begin{figure}[tb]
    \centering
    \includegraphics[width=0.95\linewidth]{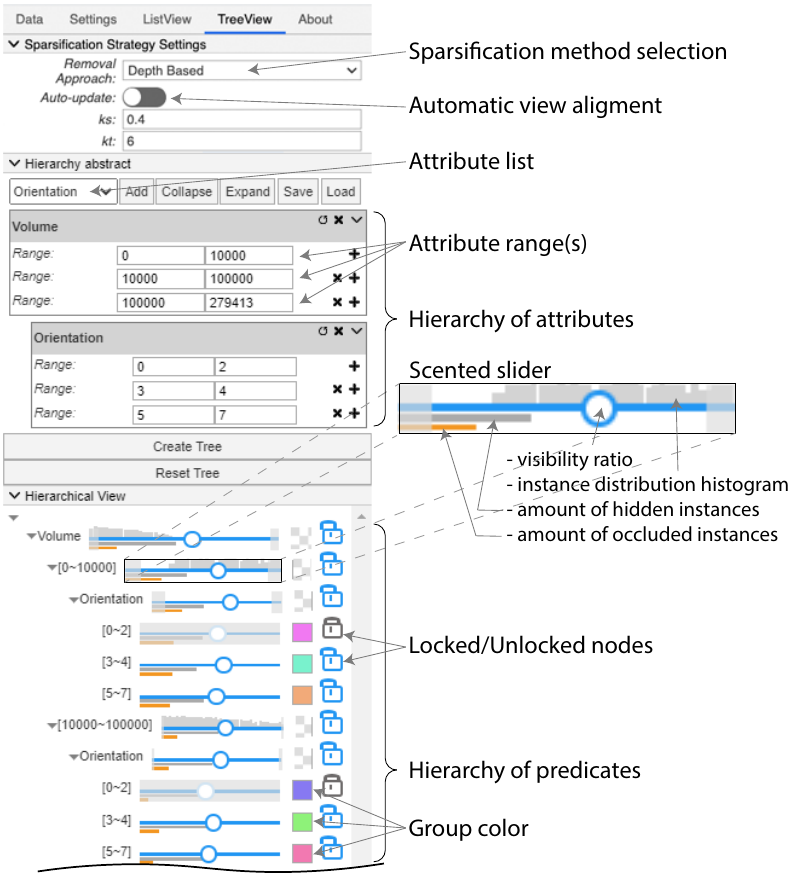}
    \caption{User interface for instance grouping with an example hierarchy. The user built an abstract hierarchy from two attributes: volume and orientation (three value ranges each). The hierarchy with the sparsification and coloring controls for each group is presented below.}
    \label{fig:tree-gui-sample}
\end{figure}

The user interface for group membership predicate manipulation is illustrated in \cref{fig:tree-gui-sample}. First, the user builds a hierarchy of attributes and defines a set of value ranges for each attribute. This form is expanded to obtain a copy of the value ranges of the children for every value range of the parent. After expansion, each path from the root to a leaf node represents a single predicate, where a single value range of a single attribute is evaluated on each level of the hierarchy.

Once the hierarchy is built, colors and visibility ratios can be applied to the resulting groups. Initially, each group is assigned a color with a random hue based on the golden ratio sequence~\cite{Schretter2012}. The user can interactively sparsify the groups by dragging their corresponding visibility ratio sliders. Sliders are assigned to all nodes of the hierarchy so that the user can change the visibility ratios of several groups simultaneously. When the user changes the visibility ratio of a group, the visibility ratios of the subgroups are updated accordingly, and vice versa, so that the visibility ratio of the parent group remains the weighted average of the visibility ratios of the subgroups based on the number of member instances. A group can also be locked so that it is not affected by such cascaded updates. Following the scented widget paradigm~\cite{Willett2007}, the scented slider for every group also depicts a histogram of the values of the corresponding attribute so that the user has a rough idea of the distribution of the attribute values of visible instances. The slider contains two additional tracks for the proportion of hidden and occluded instances. These are calculated in the visibility assessment step described in \cref{sec:visibility-assessment}.

\subsection{Instance sparsification}
\label{sec:instance-sparsification}

The sparsification procedure determines whether an instance should be visible or hidden. We only consider binary visibility to keep the user interface simple, although our implementation readily supports partial transparency via the generated transfer function. The sparsification procedure is designed as an extension of the voxel-based methods (e.g.,~\cite{Viola2005,Bruckner2006}) so that the importance of an instance is the average importance of its constituent voxels. First, we randomly shuffle the instances to prevent any correlations between their initial order and spatial distribution. When the sparsification begins, every voxel is assigned an importance value, which is aggregated over the instances. Afterward, the instances in each group are sorted based on importance, and those with the lowest importance are hidden. The number of instances to be hidden is determined by the visibility ratio of the group. The function that assigns importance to a voxel is called the \emph{sparsification function}. We propose three sparsification functions, each serving a different purpose in the visualization:
\begin{itemize}[noitemsep]
    \item \emph{uniform}, defined as follows:
    \begin{equation*}
        p_u(\pos) = 1,
    \end{equation*}
    which assigns a uniform importance and is used to sparsify the volume without changing the data distribution pattern;
    \item \emph{depth-based}, defined as follows:
    \begin{equation*}
        p_d(\pos) = \norm{\pos - \eye},
    \end{equation*}
    which assigns importance based on the distance from the camera $\eye$ and is used to create a peeling effect;
    \item \emph{context-preserving}, based on the context-preserving model~\cite{Bruckner2006} defined as follows:
    \begin{equation*}
        p_c(\pos) = \norm{\nabla V(\pos)}^{(\kappa_t \cdot s(\pos) \cdot p_d(\pos))^{\kappa_s}}.
    \end{equation*}
    See below for a detailed explanation of the different quantities.
\end{itemize}

Uniform sparsification keeps the spatial distribution of the instances unchanged, whereas depth-based sparsification reveals the internals of the volume similar to a cutaway plane, but without cutting through the instances (see \cref{fig:montage-fibers} and the online supplemental material\footnote{\label{note:supplemental}\url{https://github.com/UL-FRI-LGM/vpt-conductor/raw/master/supplemental.pdf}}). To balance the two effects, we adapted the context-preserving model~\cite{Bruckner2006}, which exhibits a similar cutaway plane functionality while allowing us to adjust the sharpness and depth of the cutaway. In the original paper, the context-preserving method was used to reduce the opacity of the less critical samples, whereas we instead use it to compute the importance of an entire instance. Conceptually, the model places a virtual light into the scene and assigns less importance to instances that receive considerable amount of light, are located closer to the camera, and are internally more homogeneous. The virtual light acts as a melting source that more strongly affects the instances with a smaller projected area toward the light. This outcome is a direct consequence of the shading factor $s(\pos)$, for which we use the Blinn-Phong shading model. To keep the user interface simple, we place the virtual light at the same position as the camera, though it could be placed anywhere. Furthermore, the parameter $\kappa_t$ controls the depth of the cutaway plane, where higher values correspond to deeper cuts, while the parameter $\kappa_s$ controls the sharpness of the cut where higher values result in a sharper cut. When $\kappa_t$ is zero, this function reduces to uniform sparsification. The gradient magnitude $\norm{\nabla V(\pos)}$ acts as an indicator of homogeneity of the data, so a more heterogeneous instance is regarded as more important.

Each of the presented sparsification functions has specific uses, and the power of this approach comes from the ability to combine them. When the density of the instances is extremely high, the user might choose to first bring it down to a reasonable level with uniform sparsification, and then use a more sophisticated approach, such as depth-based or context-preserving sparsification, as these are view-dependent. To achieve this, our method tracks which instances have already been hidden and prioritizes them during sorting. Consequently, the user can change the sparsification function on the fly and layer the results without any additional controls in the user interface. Refer to the supplementary video demonstrating this feature.

\subsection{Visibility mask}
\label{sec:visibility-mask}

\begin{figure}[tb]
	\centering
	\savebox{\largestimage}{\includegraphics[width=0.55\linewidth]{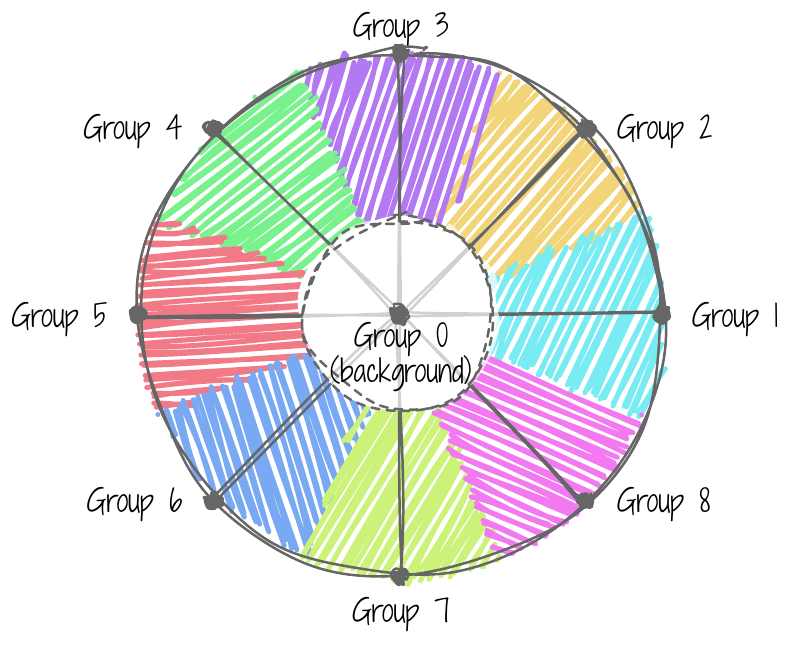}}%
    \usebox{\largestimage}
    \raisebox{\dimexpr.5\ht\largestimage-.5\height}{%
        \includegraphics[width=.35\linewidth]{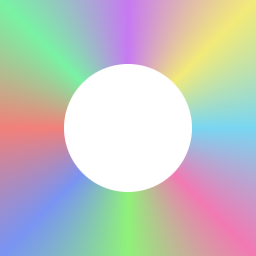}}
    \caption{Visibility mask values are mapped to the transfer function in a circular pattern. Straight lines represent interpolation paths. Paths between the background and individual groups do not intersect each other.}
    \label{fig:tfstar}
\end{figure}

To render the groups of instances, we introduce an intermediate representation called the \emph{visibility mask} that encodes the visibility of instances, their group membership, and the color and transparency for rendering. Formally, the visibility mask is a map from $\mathbb{R}^3$ to $\mathbb{R}^2$, constructed so that the voxels belonging to different groups (including the background) map to different 2D vectors. We store the visibility mask as a volume on the GPU. We generate a corresponding transfer function and use the 2D vectors from the visibility mask as texture coordinates for the transfer function, matching traditional post-classification volume rendering. Consequently, we can use any existing volume rendering algorithm to render the visibility mask. In general, the 2D vectors in the visibility mask may be arbitrary, as long as different groups of instances map to different vectors. To make the best possible use of the available transfer function space, we decided to arrange the values in a circular pattern, as illustrated in \cref{fig:tfstar}. The background is mapped to the center of the transfer function, whereas the individual groups are spread uniformly around the inscribed circle. These locations are denoted as $\maskValue$ and are defined as follows:
\begin{gather*}
    \maskValue(k) =
    \begin{cases}
        (\frac{1}{2}, \frac{1}{2}), & k = 0, \\
        (\frac{1}{2}, \frac{1}{2}) + (\frac{1}{2}\cos\phi, \frac{1}{2}\sin\phi), & k > 0,
    \end{cases} \\
    \phi = \frac{2 \pi (k - 1)}{N},
\end{gather*}
where $k$ is an index of a group (with zero being the background), and $N$ is the total number of groups. This arrangement ensures an uninterrupted interpolation path between the mask values of each group and the background, preventing any classification-related artifacts at instance boundaries, which would not be possible with a 1D transfer function. Intergroup boundaries may still cause slight rendering artifacts; however, they were barely noticeable in the use cases, and correct treatment would necessitate a far more complex solution, such as the one by Al-Thelaya \etal~\cite{Thelaya2021}.

\begin{figure*}[tb]
    \centering
    \includegraphics[width=0.9\linewidth]{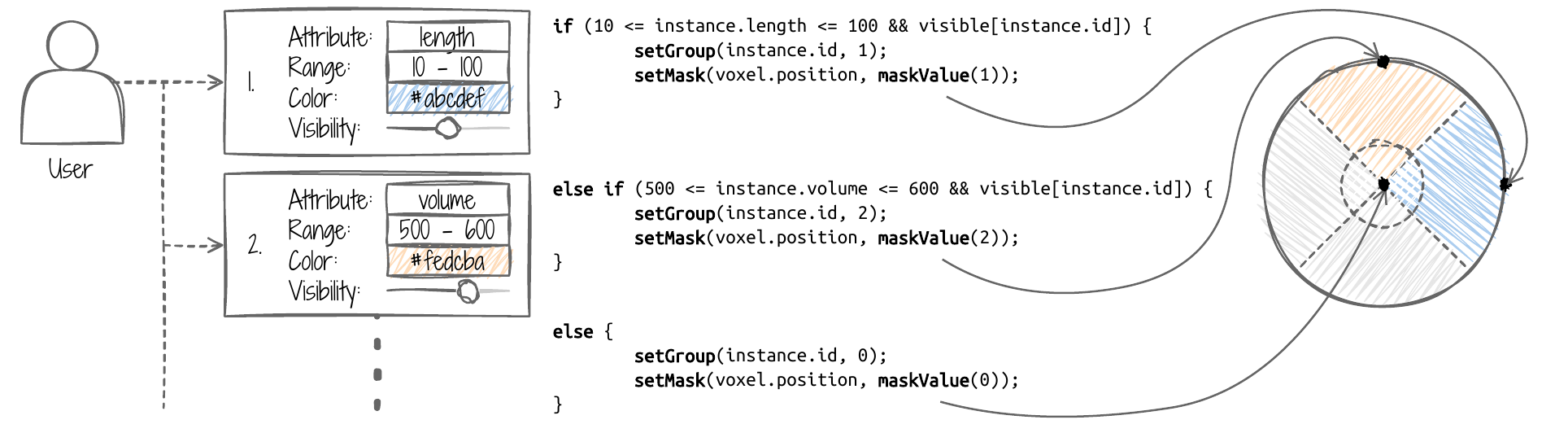}
    \caption{Shader generation from group membership predicates. The predicate list is converted to a sequence of if-else statements that assign a mask value to the voxels of the segmentation volume. The \texttt{visible} array is the result of sparsification and indicates whether an instance should be shown or hidden. The transfer function corresponding to the mask values is generated.}
    \label{fig:predicates}
\end{figure*}

We compute the visibility mask on the GPU with a procedurally generated compute shader, which makes our approach fast, general, and easily extensible. The list of group membership predicates translates into a sequence of if statements that takes an instance with its attributes as input and outputs the mask value of the corresponding group (see \cref{fig:predicates} and the algorithm in the online supplemental material\cref{note:supplemental}). The if statements are augmented so that sparsification is included by assigning the mask value of the background to the hidden instances. The compute shader is run for all voxels to generate the visibility mask. The corresponding transfer function is generated by wrapping a $1 \times N$ strip of pixels around the transparent center of the transfer function (see \cref{fig:tfstar}).

The motivation behind this design lies in the inability to interpolate the integer labels of the segmentation volume. One strategy to overcome this is to use nearest neighbor sampling, but this would result in a blocky volume and low-quality image. This problem was also recognized by Al-Thelaya \etal in a recent contribution~\cite{Thelaya2021}, but their solution involves considerable processing time and a specialized rendering algorithm, as the solution is primarily targeted at data analysis, not rendering. By contrast, the visibility mask is much simpler and orders of magnitude faster to compute and does not necessitate a specialized rendering algorithm. Additionally, we can leverage hardware interpolation of the 2D vectors from the visibility mask and use post-classification during rendering. Thus, we preserve the high-frequency details in the volume and retain sufficient rendering quality.

\subsection{Rendering and blending}
\label{sec:rendering-and-blending}

The volume conductor is independent of the rendering method. We chose to demonstrate its functionality using two rendering methods: directional occlusion shading~\cite{Schott2009} because it can simulate direct illumination at interactive frame rates, and path tracing because it produces physically realistic results (refer to the online supplemental material\cref{note:supplemental} for a visual comparison). We augmented the sampling procedure with blending between the visibility mask and raw data volume to allow the user to see the internals of the instances while using the sparsification functionality. In addition to visibility mask $V_\mathrm{mask}$ and its corresponding transfer function $TF_\mathrm{mask}$, we have raw data volume $V_\mathrm{raw}$ and its transfer function $TF_\mathrm{raw}$ defined by the user. Both are sampled to obtain color $C$ and opacity $A$ from position $\pos$ within the volume:
\begin{align*}
    (C_\mathrm{mask},\ A_\mathrm{mask}) &= TF_\mathrm{mask}(V_\mathrm{mask}(\pos)), \\
    (C_\mathrm{raw},\ A_\mathrm{raw}) &= TF_\mathrm{raw}(V_\mathrm{raw}(\pos)).
\end{align*}
We linearly blend the colors with interpolation weight $w_\mathrm{color}$:
\begin{equation*}
    C_\mathrm{final} = (1 - w_\mathrm{color}) C_\mathrm{mask} + w_\mathrm{color} C_\mathrm{raw}.
\end{equation*}
We also subject the opacity of the raw data to sparsification; therefore, it is treated slightly differently than the colors. First, we transfer the opacity from the visibility mask to the raw data volume with interpolation weight $w_\mathrm{transfer}$:
\begin{equation*}
    A_\mathrm{transfer} = (1 - w_\mathrm{transfer}) A_\mathrm{mask} + w_\mathrm{transfer} A_\mathrm{mask} A_\mathrm{raw}.
\end{equation*}
Then, we compute the final opacity as a linear interpolation between the transferred opacity and raw data opacity with interpolation weight $w_\mathrm{alpha}$:
\begin{equation*}
    A_\mathrm{final} = (1 - w_\mathrm{alpha}) A_\mathrm{transfer} + w_\mathrm{alpha} A_\mathrm{raw}.
\end{equation*}
The user sets all interpolation weights in the user interface. When $w_\mathrm{color}$ and $w_\mathrm{alpha}$ are both 1, this reduces to raw data volume rendering. When both weights and $w_\mathrm{transfer}$ are 0, only the visibility mask defines the output.

\subsection{Visibility assessment}
\label{sec:visibility-assessment}

Due to the occlusion and perspective projection, the density of the instances after sparsification often does not precisely reflect the visibility ratio set by the user. We measure the actual visibility ratio of the instances as observed from the camera and inform the user about the number of occluded instances. We update an ID frame buffer during rendering, which holds the ID of the nearest visible instance for every pixel and the group to which it belongs. We measure the number of visible instances in each group by counting the unique instance IDs from the group present in the ID frame buffer. We know the total number of instances in each group and the number of hidden instances; thus, we also know how many instances from that group are occluded. This information is presented to the user through an additional track under the scented slider (see \cref{fig:tree-gui-sample}).

\section{Implementation Details}
\label{sec:implementation-details}

With the focus on making the method interactive, we implemented the computationally intensive parts on the GPU and integrated them into the VPT framework~\cite{Lesar2018} using WebGL 2.0 Compute\footnote{\url{https://github.com/UL-FRI-LGM/vpt-conductor}}. We also retained the ability to perform hardware sampling and interpolation to ensure the method remains independent of the rendering technique. We store the segmentation volume on the GPU as a 3D texture of 32-bit unsigned integers (format \texttt{R32UI}) and their respective attributes in a shader storage buffer object (SSBO). The layout and format of the data in the SSBO are such that it directly maps to an array of \texttt{structs} in a shader, where each \texttt{struct} holds the attributes of a single instance. Thus, each integer value from the segmentation volume acts as an index for accessing the attributes of the corresponding instance. As the layout of the \texttt{struct} varies between datasets, it must be supplied along with the volumes and instance attributes. We use a simple JSON file with the attribute names and types, which is enough to generate the struct definition in GLSL. We access the instance attributes in the procedurally generated compute shader when generating the visibility mask. As displayed in \cref{fig:predicates}, the group membership predicates translate directly into a sequence of GLSL if-else statements, which assign a mask value to every voxel in the visibility mask, similar to the approach by Schulte zu Berge \etal~\cite{SchultezuBerge2014}, although our predicates are not used during rendering but instead are employed to generate the visibility mask. We store the resulting visibility mask on the GPU as a 3D texture of 2D 8-bit normalized integer vectors. We use the format RGBA8 because it is one of the few that can be written to using a WebGL 2.0 compute shader (in contrast to RG8, which would be more appropriate), and it allows hardware interpolation. The compute shader is regenerated after every change in the group membership predicates. The shader execution is independent between voxels, so no communication is needed between the workgroups, and the size of the workgroups may be arbitrary. In our implementation, we chose $16 \times 16 \times 1$ because it maps reasonably well to most modern hardware.

\section{Results}
\label{sec:results}

\subsection{Performance evaluation}

\begin{figure}[tb]
	\centering
	\includegraphics[width=\linewidth]{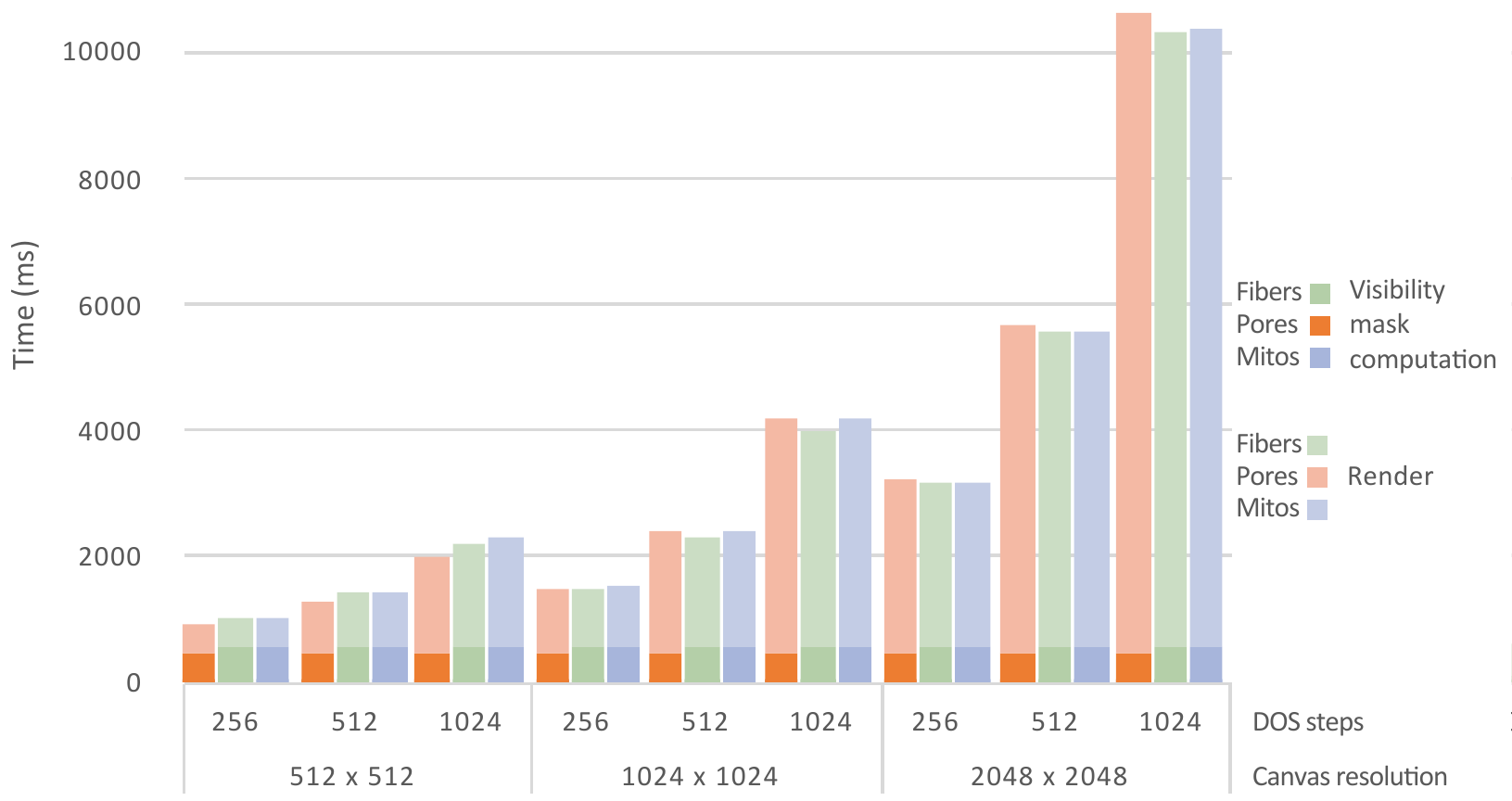}
	\caption{Performance evaluation for three datasets: fibers, pores, and mitochondria on a laptop computer with integrated graphics. The times to compute the visibility mask and render the image are listed. Linearization and shader recompilation times are negligible and not depicted in the graph.}
	\label{fig:results-graph}
\end{figure}

We evaluated the volume conductor on three computers: a laptop with Intel HD Graphics 530 integrated graphics with 20 GB of shared RAM, a desktop computer with an Nvidia GeForce GTX 1060 graphics card with 6 GB of RAM, and a professional workstation with an Nvidia Quadro RTX 8000 graphics card with 48 GB of RAM. Three datasets were used: fibers, pores, and mitochondria, presented in \cref{subsec:use-cases}. We measured the time of execution of the following three steps:
\begin{enumerate}[noitemsep]
    \item linearization of the predicate hierarchy and shader recompilation,
    \item visibility mask computation, and
    \item rendering with directional occlusion shading.
\end{enumerate}
The measurements were executed 10 times (10 linearizations, 10 visibility mask computations, and 10 rendered frames), and the average time per step was recorded. We used the Google Chrome web browser version 87. The evaluation was performed for a nontrivial predicate hierarchy with two layers and five ranges for each layer. The camera was configured so that the volume filled the screen. This configuration is essential because the rendering time is highly view-dependent. All measurements are gathered in the appendix, and the measurements for the integrated graphics are presented graphically in \cref{fig:results-graph}.

The performance evaluation demonstrates that the presented approach remains interactive on platforms with varying capabilities. \cref{fig:results-graph} and the related graphs in the supplemental material\cref{note:supplemental} show much better performance of the dedicated graphics hardware. The visibility mask computational time is proportional to the volume size, and the rendering time is proportional to the frame buffer resolution, both of which conform to expectations. The measurements reveal one peculiarity in the shader rebuild time, which is substantially smaller for integrated graphics than dedicated graphics hardware. This discrepancy is likely caused by the additional communication time between the CPU and dedicated GPU. Due to the nonoptimized implementation, the most computationally intensive part of the volume conductor is the rendering process. Keep in mind that the rendering method can be easily swapped. The visibility mask computation time is more relevant to the evaluation of the method than rendering time, and is low enough for interactive use even on commodity hardware. In fact, in a realistic scenario, the visibility mask is recomputed only when the user modifies the instance groups or the visibility ratio of a group, but not during rendering.

\subsection{Use cases}
\label{subsec:use-cases}

We tested the volume conductor on two domains: fiber-reinforced polymers in the field of material science and intracellular organelles in the field of microbiology. Volumetric images from both domains are crowded and, therefore, a perfect fit for the volume conductor. We collaborated with experts from these two fields and instructed them to use the volume conductor in their daily workflow. After a few days, we gathered their feedback and asked them to state the advantages and disadvantages of the method.

\subsubsection{Material Science}

\begin{figure}
    \centering
    \includegraphics[width=\linewidth]{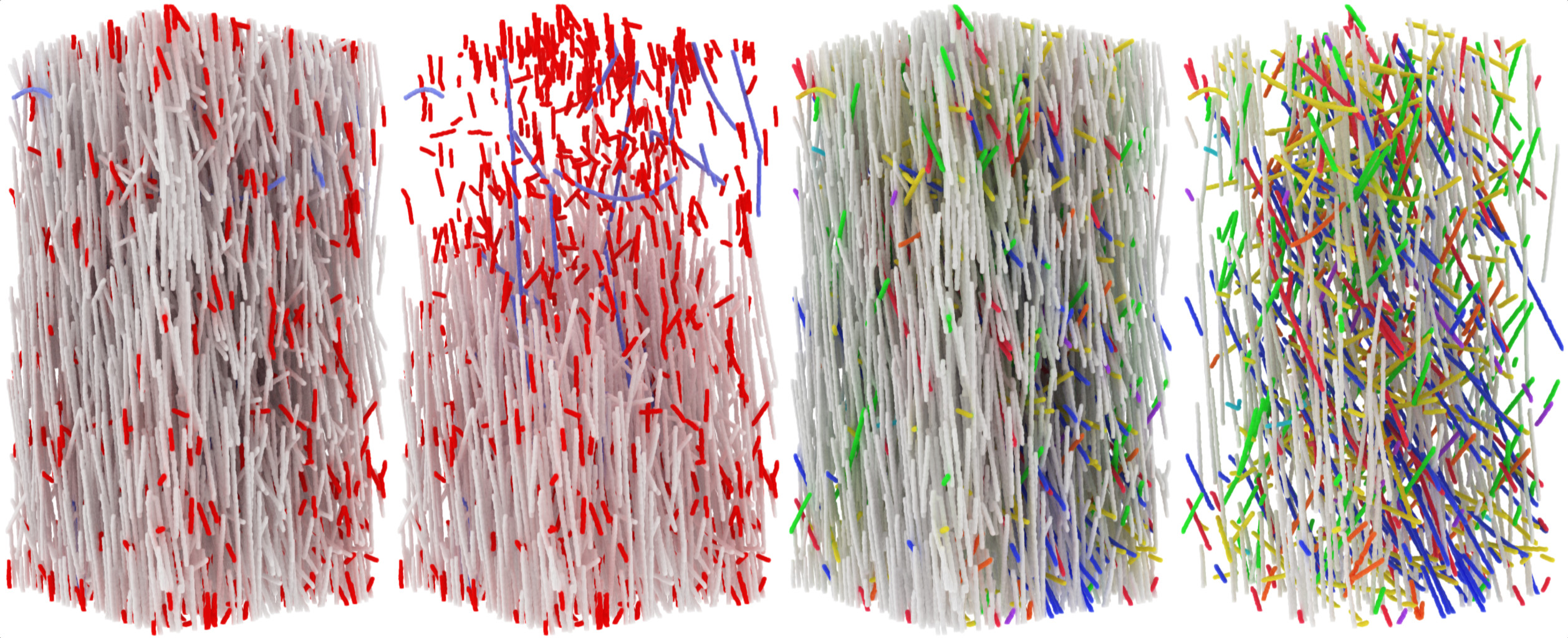}
    \caption{Fiber use case. From left to right: colorization of short (red) and bent (blue) fibers, the same as previous but with 50~\% of the remaining fibers hidden by the depth from above, and colorization by orientation, the same as previous but with 80~\% of the vertical fibers hidden with the context-preserving model.}
    \label{fig:montage-fibers}
\end{figure}

Fiber-reinforced polymers are in high industrial demand due to their strength, durability, elasticity, and low weight, and the demand is steadily growing~\cite{Heinzl2017}. Their physical properties are directly related to the distribution and density of internal structures, such as fibers, inclusions, and pores, and their properties, such as length, volume, and orientation. Therefore, a tool that can isolate and visualize structures with specific properties is crucial for scientists to analyze and improve the materials. In this domain, a typical imaging technique is the 3D X-ray CT. DVR renderings of the resulting volumes are often difficult or even impossible to interpret due to crowdedness. Specialized analysis tools are available, such as Feature Scout~\cite{Weissenbock2014}.

We considered two datasets: (1) a $400 \times 401 \times 800$ volume containing 3828 glass fiber instances with 18 attributes each, and (2) a $512 \times 512 \times 512$ volume containing 6888 instances of pores between carbon fibers with 41 attributes each. The use cases were provided by the University of Applied Sciences, Upper Austria, a research associate with three years of experience and a senior researcher with more than 15 years of experience in material science data visualization. The experts compared the use of Feature Scout and volume conductor on two use cases commonly encountered in their everyday work. They were also involved in the design process for both systems and had good insight into their functionalities.

\paragraph{Use Case 1 -- Fiber analysis} In glass-fiber-reinforced materials, fiber characteristics must be analyzed, and the spatial distribution of the fibers with specific properties must be determined. In a strong material, the fibers are uniformly oriented and uniformly distributed across the volume. Short and bent fibers that do not contribute to the strength of the material can be quickly identified by colorization (\cref{fig:montage-fibers}, two left images). The interactive sparsification feature of the volume conductor can be used to obtain a general overview of the directional distribution of the fibers (\cref{fig:montage-fibers}, right two images).

\begin{figure}
    \centering
    \includegraphics[width=\linewidth]{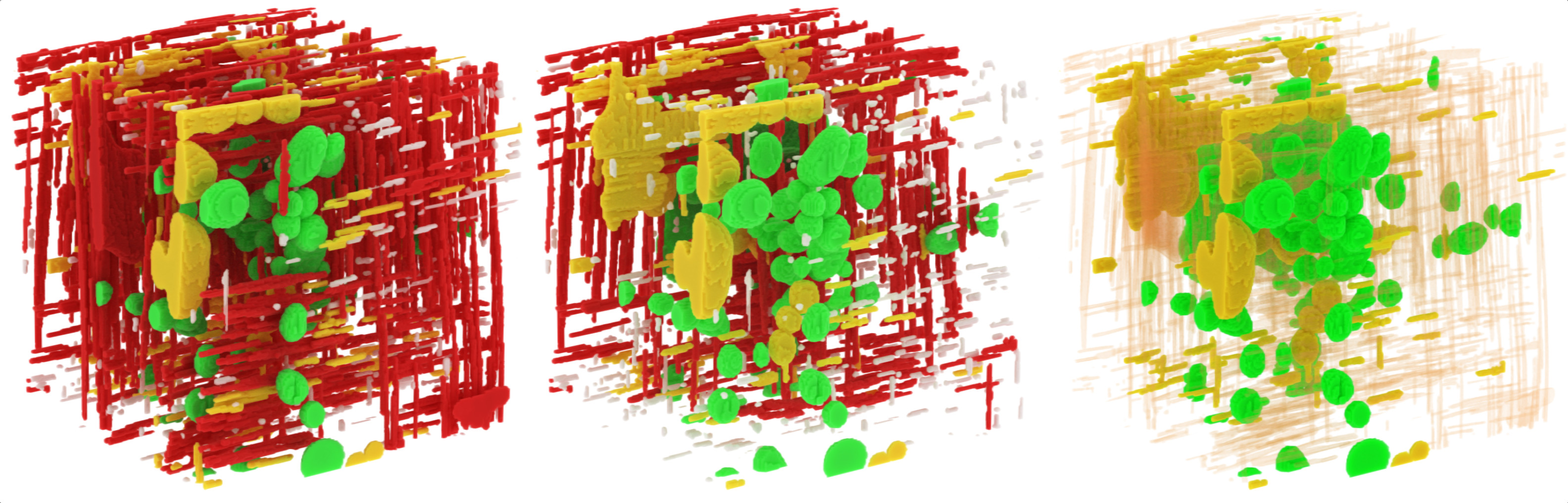}
    \caption{Use case with pores. From left to right: colorization by roundedness (needle-shaped pores are red), the same as previous but with 50~\% of the needle-shaped pores hidden by depth, and ghosting.}
    \label{fig:montage-pores}
\end{figure}

\paragraph{Use Case 2 -- Pore analysis} In carbon-fiber-reinforced polymers, the existing pores inside the material must be examined, especially their shape and spatial distribution. For example, needle-shaped pores have a higher potential for crack initiation than elliptical or spherical pores. With both the volume conductor and Feature Scout, an appropriate categorization can be constructed to obtain a visual overview of the needle-shaped pores by categorizing the instances according to shape (\cref{fig:montage-pores}, two left images). In contrast to Feature Scout, the volume conductor provides interactive sparsification for a better overview of the pore distribution.

After using the volume conductor, the material science experts came to the following conclusions:
\begin{itemize}[noitemsep]
    \item Uniform sparsification primarily helps when sparsifying the crowded volume of fibers while maintaining the distribution of the instances throughout the volume.

    \item View-aligned sparsification functions (depth-based and context-preserving functions) are beneficial during exploration of the internals of the volume (\cref{fig:montage-pores}, right two images). The ability to layer them on top of uniform sparsification is a very powerful feature.

    \item On-the-fly changes to the membership predicates accelerate the exploration process because the user can add additional predicates and refine the visualization at will, which is much more tedious in Feature Scout.

    \item The ability to select individual instances directly in the rendering and inspect their attributes is missing.
\end{itemize}
The experts claim that sparsification is an invaluable tool for data exploration in both use cases because it allows them to immediately observe the distribution of fibers and pores with specific properties. The experts stated that a featureful tool, such as Feature Scout, could benefit from the volume conductor and help them in their everyday workflow.

\subsubsection{Cell biology}

\begin{figure}
    \centering
    \includegraphics[width=\linewidth]{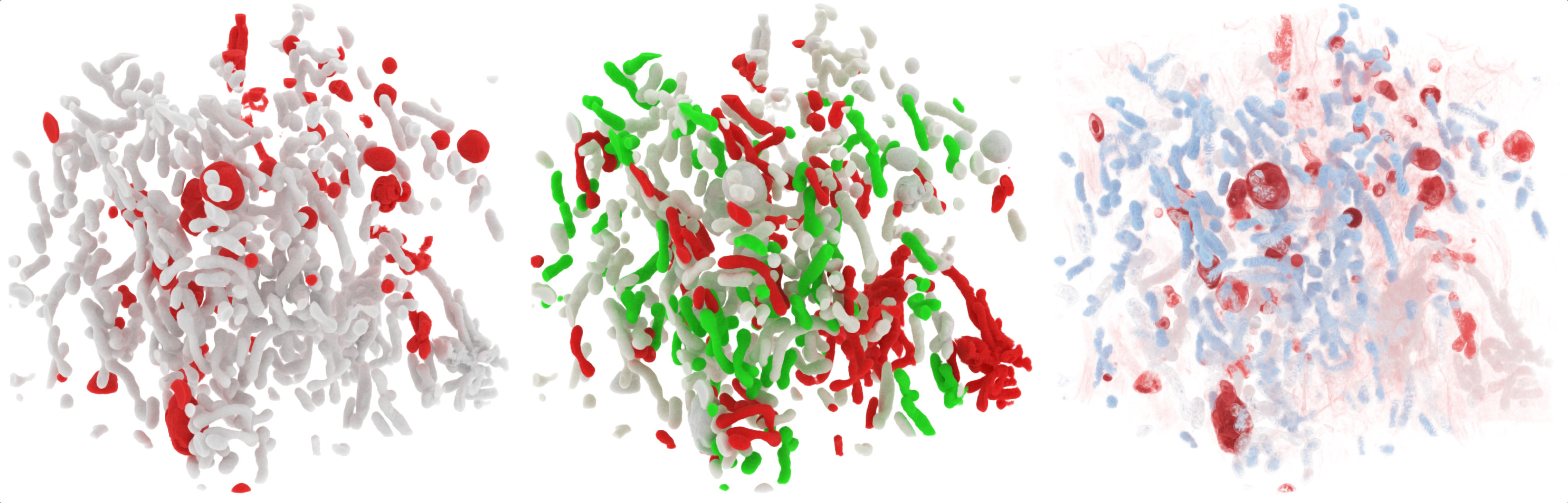}
    \caption{Mitochondria use case. From left to right: colorization by organelle type (mitochondria are red), colorization of branched (red) and thinned (green) mitochondria, blending with raw data.}
    \label{fig:montage-mitos}
\end{figure}

We tested the volume conductor on a 3D microscopy sample of a cell inside a mouse bladder. The volume was provided by the Institute of Cell Biology of the University of Ljubljana. Microbiologists at the institute study the distribution, density, and shape of intracellular organelles to explore and understand various cellular processes. In one specific case, they were interested in the mitochondria and endolysosomes regarding their size, shape, curvature, and possible branchings and narrowings (\cref{fig:montage-mitos}, middle image). Considering that thousands of such organelles may exist even in a small subsection of a cell, scientists need a tool, such as the volume conductor, to visualize and analyze organelles and their properties.

The $1280 \times 1024 \times 1024$ microscopy sample was segmented, and the features were extracted (see~\cite{ZerovnikMekuc2020} for details), yielding 3051 instances with 21 attributes each. The use case was given by the Institute of Cell Biology at the Faculty of Medicine of the University of Ljubljana: an assistant professor with 16 years of experience and a professor with more than 30 years of experience. They are both familiar with the software used in their fields, such as ImageJ and 3D Slicer. They tested the volume conductor in a one-day test run under the supervision of one of the authors. The professors compared the volume conductor with the software stack they use in their regular work.

\paragraph{Use Case 3 -- Mitochondria analysis} For analysis and exploration of the volumetric microbiological data obtained using modern microscopy techniques (e.g., focused ion beam scanning electron microscopy), researchers still primarily employ slice-based visualization tools (e.g., ImageJ) or regular DVR tools (e.g., 3D Slicer). The tasks, such as examining the distribution of intracellular organelles, are very time-consuming and demanding. The volume conductor facilitates these tasks considerably and enables users to perform complex analyses. For example, the shapes, quantity, orientation, and distribution of mitochondria can be studied using the grouping and sparsification functionality of the volume conductor. Additionally, microbiologists can view the segmented intracellular structures with raw volumetric data, allowing them to visualize the internals of the organelles in a sparsified environment (\cref{fig:montage-mitos}, right image).

After using the volume conductor, microbiology experts came to the following conclusions:
\begin{itemize}[noitemsep]
    \item It is beneficial to sparsify the volume and colorize the instances with specific properties. This method allows new possibilities both in the visualization and exploration processes.

    \item The set of sparsification functions enables easier volume exploration and provides more refined control over what to display and what to hide.

    \item A joint visualization of raw and segmented data offers the user a view of the interior of the segmented structures without the surrounding clutter.
    
    \item A disadvantage is that the user must provide the segmentation volume and instance attributes, which are not always available or easily computed or obtained.
    
    \item It would be beneficial if such a visualization tool was coupled with a specialized analysis tool. Specifically, selecting individual instances by clicking on them in the rendering and observing their properties would be a very powerful feature.
\end{itemize}

\section{Discussion}
\label{sec:discussion}

The feedback from both expert groups reveals that the interactive sparsification functionality is the most valuable feature of the volume conductor. The volume conductor provides a means for acquiring a holistic view of the data before a detailed analysis. It shifts the focus from a low-level view of the raw data to a high-level view of the structures and their properties, and the user interface is designed to support this view. For the material science experts, the main benefit of the volume conductor is sparsification and its independence from the rendering method, which makes the volume conductor particularly easy to integrate into existing tools. The volume conductor enhances them with sparsification capabilities, as long as at least a crude instance segmentation is available. According to the microbiology experts, this reliance on segmentation availability is one of the main disadvantages of the volume conductor. Obtaining an accurate instance segmentation is a tedious process if automatic tools are not available, and without it, the volume conductor is inoperative. However, a crude segmentation is often enough to reap the benefits of the volume conductor. In contrast to fiber-reinforced polymers, the microbiological use case revealed a strong need for a joint display of segmented and raw data.

Experts from both groups pointed out the ease of use of the volume conductor due to the user interface designed specifically for this purpose. However, when using a joint display of segmented and raw data, the user must still design an appropriate transfer function for the raw data. This necessity results in a usability gap between the two rendering modes. The user can become lost even in the easy-to-use user interface when presented with a wide array of possibilities for the attribute hierarchy, and the confusion only intensifies with a long list of instance attributes. It is not unusual to compute more than 20 different features for a single instance. With machine learning, this number of features is easily surpassed by many orders of magnitude. Machine learning could eventually be used for automatic instance grouping, colorization, and sparsification to shield the user from the numerous resulting possibilities for visualization. However, this is out of the scope of this paper and will be considered for future work.

\section{Conclusion}
\label{sec:conclusion}

We demonstrated how the volume conductor can be beneficial for exploring crowded volumes compared to regular DVR. With the proposed method, the user can group the instances based on simple predicates or a hierarchy of predicates and interactively adjust their density to reveal more information about the structure of the volume and the distribution of instances. We achieved this by separating instance grouping, visibility mask computation, transfer function generation, and rendering. The resulting method is general and easily extensible due to the procedurally generated compute shader. We applied the technique to two domains, in which the experts reported improvements in their workflow and exploration process. In the future, we intend to improve the volume conductor with single or multiple instance selections in the cases where the user aims to make fine adjustments to the visibility of specific instances. We also plan a method for automatic or user-guided instance grouping to make the exploration process even faster and simpler. We believe the method will prove invaluable for interactive data exploration in many research fields where crowded volumetric environments are in use.



\section*{Acknowledgments}
The authors would like to thank Julia Maurer from the University of Applied Sciences, Upper Austria, for providing carbon-fiber-reinforced polymer data and the presented use cases; Samo Hudoklin and Rok Romih from the University of Ljubljana for providing cell structure data and the presented use cases. The research was partially supported by the King Abdullah University of Science and Technology -- award number BAS/1/1680-01-01, partially by funding from the Austrian Research Promotion Agency (FFG) within the program line ``TAKE OFF,'' FFG grant no. 874540 ``BeyondInspection,'' and by research subsidies granted by the government of Upper Austria during the ``X-Pro'' project.

\bibliographystyle{spmpsci}
\bibliography{bibliography}

\end{document}